\documentclass[aps,preprint]{revtex4}
\usepackage{amsmath,epsfig,graphicx,amssymb}

\begin{document}
\title{TDDFT WITH SKYRME FORCES: EFFECT OF TIME-ODD DENSITIES ON ELECTRIC GIANT
RESONANCES}
\author{V.O. Nesterenko$^{1}$, W. Kleinig $^{1,2}$, J. Kvasil $^{3}$, P. Vesely $^{3}$,
and P.-G. Reinhard $^{4}$}
\address{$^{1}$ Bogoliubov Laboratory of Theoretical Physics,
Joint Institute for Nuclear Research, Dubna, Moscow region, 141980, Russia}
\address{$^{2}$  Technische Univirsitat Dresden, Inst. f\"ur Analysis,
   D-01062, Dresden, Germany}
\address{$^{3}$ Institute of Particle and Nuclear Physics, Charles University, CZ-18000 Praha 8,
Czech Republic}
\address{$^{4}$Institut f\"ur Theoretische Physik II,
Universit\"at Erlangen, D-91058, Erlangen, Germany}

\begin{abstract}
Time-odd densities and their effect on electric giant resonances
are investigated within the self-consistent separable
random-phase-approximation (SRPA) model
for various Skyrme forces (SkT6, SkO, SkM*, SIII, SGII, SLy4, SLy6, SkI3).
Time-odd densities restore Galilean invariance of the Skyrme functional,
violated by the effective-mass and spin-orbital terms.
In even-even nuclei these densities do
not contribute to the ground state but can affect the dynamics. As a
particular case, we explore the role of the current density in description of
isovector E1 and isoscalar E2 giant resonances in a chain of Nd
spherical and deformed isotopes  with A=134-158. Relation of the
current to the effective masses and relevant parameters of the Skyrme
functional is analyzed. It is shown that current contribution to E1 and E2
resonances is generally essential and fully determined by the values
and signs of the isovector and isoscalar effective-mass parameters of
the force. The contribution is the same for all the isotope
chain, i.e. for both standard and exotic nuclei.
\end{abstract}

\maketitle

\section{Introduction}

Skyrme forces \cite{Skyrme,Vau} are widely used for exploration of the
ground state and excitations in atomic nuclei. In particular, last years
big effort was done in application of these forces to exotic nuclei,
nuclear matter and
astrophysical problems (for recent reviews see \cite{Ben_rew,Stone_rew}).
However, in spite of the impressive progress, Skyrme forces
still suffer some serious troubles. For example, there are too many Skyrme
parametrizations and the forces do require a unification.
Further, being successful in description of the nuclear ground states,
Skyrme forces often fail in a comprehensive treatment of nuclear
dynamics (e.g. no one force can describe  equally well both isoscalar and
isovector  modes, see \cite{nest_06} for recent discussion).
So, the further upgrade of the forces is necessary and nuclear dynamics should
play here an important role.

In this connection, we propose here the analysis of time-odd densities
in the Skyrme functional and their role in description of electric giant
resonances (GR). Time-odd densities (current $\vec j$, spin $\vec s$, and
vector kinetic-energy $\vec T$) are known to restore Galilean invariance of
the Skyrme functional, violated by velocity-dependent time-even densities
(kinetic-energy $\tau$ and spin-orbital $\vec\Im$) \cite{Engel_75,Dob}.
Therefore time-odd densities are related to important features of
the functional (effective masses and spin-orbital interaction)
and might play an essential role in its upgrade. In this paper, we discuss
time-odd densities in general and, as a particular case, scrutinize the influence
of the current density $\vec j$ on the isovector E1(T=1) and isoscalar E2(T=0)
GR in a chain of Nd isotopes with A=134-158. As compared with the previous
studies \cite{nest_06,nest_07}, much more nuclei and forces are involved to make
the exploration indeed systematic. We will establish important relations of the
GR properties to specific parameters and terms of Skyrme forces and, on these
grounds, propose a tentative classification of the forces.

The calculations are performed within the  self-consistent
random-phase-approximation (SRPA) method with factorized Skyrme forces
\cite{nest_06,Kva_miori_01,nest_02,ne_les_houches_04,prep_05}. SRPA
covers both spherical \cite{nest_02} and deformed \cite{nest_06,prep_05}
nuclei. Factorization minimizes  the computational
effort minimized and makes possible the systematic exploration.

\section{Time-odd densities}
\subsection{General properties}

The Skyrme functional \cite{Skyrme} in the form
\cite{Vau,Engel_75,Dob,Re92}
\begin{equation}
   {E}  =  \int d{\vec r} ( {\cal H}_{\rm kin}
   + {\cal H}_{\rm C}(\rho_p) + {\cal H}_{\rm pair}(\chi_q)
   + {\cal H}_{\rm Sk}(\rho_q,\tau_q,
   \vec{s}_q,\vec{j}_q,\vec{\Im}_q) ) \; ,
\end{equation}
includes  kinetic, Coulomb, pairing and Skyrme terms respectively.
Expressions for the first three terms are done elsewhere
\cite{nest_02,nest_07,prep_05,Re92}.  The Skyrme part reads
\begin{eqnarray}
\label{Esky}
   {\cal H}_{\rm Sk} &= &
                 \frac{b_0}{2}  \rho^2
                  -\frac{b'_0}{2} \sum_q \rho_q^2
    -\frac{b_2}{2} \rho (\Delta \rho)
     +\frac{b'_2}{2} \sum_q \rho_q (\Delta \rho_q)
 \\
          &+&
          \frac{b_3}{3}  \rho^{\alpha +2}
          - \frac{b'_3}{3} \rho^\alpha   \sum_q\rho_q^2
     +  b_1 (\rho \tau - \vec{j}^{\; 2})
     - b'_1 \sum_q (\rho_q \tau_q - \vec{j}_q^{\; 2})
\nonumber\\
     &-&
      b_4 \left( \rho (\vec{\nabla}\vec{{\Im}})
      + \vec{s} \cdot (\vec{\nabla} \times \vec{j})\right)
     -b'_4 \sum_q \left( \rho_q(\vec{\nabla} \vec{\Im}_q)
      + \vec{s}_q \cdot (\vec{\nabla} \times \vec{j}_q)\right)
\nonumber\\
     &+&
      \tilde{b}_4\left( \vec{s}\vec{T}-\vec{\Im}^2 \right)
      + \tilde{b}'_4\sum_q \left( \vec{s}_q\vec{T}_q-\vec{\Im}_q^2
      \right) \; .
\nonumber
\end{eqnarray}
This part involves time-even (nucleon $\rho_q$, kinetic-energy $\tau_q$,
and spin-orbital $\vec \Im_q$) and time-odd (current $\vec j_q$, spin
$\vec s_q$, and vector kinetic-energy ${\vec T}_q$) densities, where $q$
denotes protons and neutrons. The total densities
(like $\vec j=\vec j_p + \vec j_n$) are given
in (\ref{Esky}) without the index.

Both time-even and time-odd densities follow from the original Skyrme
forces \cite{Skyrme}. Time-even densities contribute to both ground state and
excitations. Time-odd densities affect the excitations as well.
They also influence g.s. of odd and odd-odd nuclei but are irrelevant for g.s.
of spin-saturated even-even nuclei.

As was mentioned above, time-odd densities restore Galilean invariance of the
functional, violated by velocity-dependent time-even densities $\tau_q$ and
$\vec \Im_q$ \cite{Engel_75,Dob}. Just for this reason, time-odd densities enter
the functional only in the specific combinations with their time-even
counterparts. Hence, time-odd densities do not lead to any new parameters.
This is especially useful for even-even nuclei where Skyrme parameters,
being fixed for the g.s. with implementation of the time-even densities only,
can be further applied to nuclear dynamics involving time-odd densities
as well.

Explicit expressions for time-even and time-odd densities via single-particle
wave functions can be found elsewhere \cite{Ben_rew,nest_06,nest_02,Dob,Re92}.
It is more instructive to write down these densities in terms of the {\it basic}
scalar $\rho_q$ and vector $\vec s_q$ densities
\begin{equation}
\label{basic_dens}
\rho_q(\vec r,\vec r ') = \sum_{\sigma} \rho(\vec r\sigma, \vec r '\sigma'),
\quad
s_q^{\nu}(\vec r,\vec r ') = \sum_{\sigma\sigma'}
\rho_q(\vec r\sigma, \vec r '\sigma')\langle \sigma'|\hat\sigma_{\nu}|\sigma \rangle
\end{equation}
which form a general one-body density matrix \cite{Engel_75}
\begin{equation}\label{1b_dm}
\rho_q(\vec r\sigma ,{\vec r}'\sigma')
= \frac{1}{2}[\rho_q(\vec r,\vec r ')\delta_{\sigma,\sigma'}
+ \sum_{\nu}\langle \sigma|\hat{\sigma}_{\nu}|\sigma'\rangle  s_q^{\nu}(\vec r,\vec r ')]
\; .
\end{equation}
Here $\hat{\sigma}_{\nu}$ is a Pauli matrix and
$\sigma =\pm 1$ .
Then the densities in (\ref{Esky}) read as \cite{Engel_75}
\begin{eqnarray}
\label{bd}
\rho_q(\vec r)&=& \rho_q(\vec r,\vec r),
\qquad\qquad\qquad\qquad\quad\;
\vec s_q(\vec r) = \vec s_q(\vec r,\vec r),
\\
\label{bd_mom}
\vec j_q(\vec r)&=& \frac{1}{2i}[(\vec\nabla - \vec\nabla')
\rho_q(\vec r,\vec r')]_{\vec r = \vec r'},
\quad
\vec \Im_q^{\mu\nu}(\vec r) = \frac{1}{2i}
[(\vec\nabla_{\mu} - \vec\nabla'_{\mu}) s_q^{\nu}(\vec r,\vec r')]_{\vec r =\vec r'},
\\
\label{bd_ke}
\tau_q(\vec r)&=& [\vec\nabla \cdot \vec\nabla'
\rho_q(\vec r,\vec r')]_{\vec r = \vec r'},
\qquad\qquad\;
\vec T_q(\vec r) = [\vec\nabla \cdot \vec\nabla')
\vec s_q(\vec r,\vec r')]_{\vec r =\vec r'} \; ,
\end{eqnarray}
i.e. are reduced to the basic densities (\ref{bd}), their momenta (\ref{bd_mom})
and kinetic energies (\ref{bd_ke}). This presentation clarifies
physical sense of the densities in terms of hydrodynamics. Besides, the presentation
(\ref{bd})-(\ref{bd_ke}) can be treated as some kind of the gradient expansion
up to the second derivatives of the basic densities. Such expansion is obviously
useful for non-uniform densities, which is the case for $\vec s_q$ and $\rho_q$
at the nuclear boundary \cite{Ben_rew}.

\subsection{Current density}

Between time-odd densities, the current  is most important for electric GR.
Indeed, the spin density $\vec s_q$ is mainly relevant for magnetic modes and does
not influence electric GR \cite{nest_06,nest_07}. The density $\vec T_q$ can also
be neglected as it supplements in (\ref{Esky}) the term $\Im^2_q$ which is commonly
omitted in most of the Skyrme forces.

Contribution of $\vec j_q$ to the residual interaction is driven by the variation
\cite{nest_06,nest_02}
\begin{equation}
\frac{\delta^2 E}{\delta \vec{j}_{q'}(\vec r')\delta \vec{j}_{q}(\vec r)}=
2[-b_1+b_1'\delta_{q,q'}]\delta(\vec r' - \vec r)
\end{equation}
which is fully determined by the terms $\sim b_1, b_1'$ in the
functional (\ref{Esky}). In these terms, the current density
adjoins the values $\rho\tau$ and $\rho_q\tau_q$
responsible for the effective mass. Hence, one may expect the
correlation between effective masses and $\vec j_q$ in
GR dynamics.

To inspect this correlation, it is convenient to express the terms
$\sim b_1, b_1'$ through isoscalar and isovector densities
$\rho_0=\rho_n+\rho_p$ and $\rho_1=\rho_n-\rho_p$  (the same for
$\tau$ and $\vec j$). Then the sum of the terms with $b_1$ and
$b_1'$ is transformed to the form
\begin{equation}
B_0(\rho_0\tau_0-j^2_0)-B_1(\rho_1\tau_1-j^2_1)
\end{equation}
where isoscalar and isovector contributions are decoupled and
$B_0=b_1-b_1'/2$ and $B_1=b_1'/2$ \cite{Stone_rew}.
For the symmetric nuclear matter, the isoscalar effective mass
$m^*/m$ and the sum-rule enhancement factor $\kappa= (m_1^*/m)^{-1}-1$
(where $m_1^*/m$ is the isovector effective mass)  can be
expressed via the new parameters as
\begin{equation}\label{rel_m*}
(\frac{m^*}{m})^{-1} = 1+2\frac{m}{\hbar^2}\bar\rho B_0, \quad
\kappa = \frac{2m}{\hbar^2}\bar\rho (B_0+B_1)
\end{equation}
where $m$ is the bare nucleon mass and $\bar\rho$ is the density of
the symmetric nuclear matter. The relations (\ref{rel_m*})
are illustrated in Fig. 1 for the set of Skyrme forces used in this paper.
We see that there indeed take place correlations $B_0 \leftrightarrow m^*/m$
and $B_1 \leftrightarrow \kappa$. In the Section 4 we will show that
$\vec j_q$-impact to E2(T=0) and E1(T=1) GR is also determined
by the parameters $B_0$ and $B_1$. This will justify the relation between
$\vec j_q$ and effective masses.
\begin{figure}
\centerline{
\includegraphics[width=8cm,angle=-90]{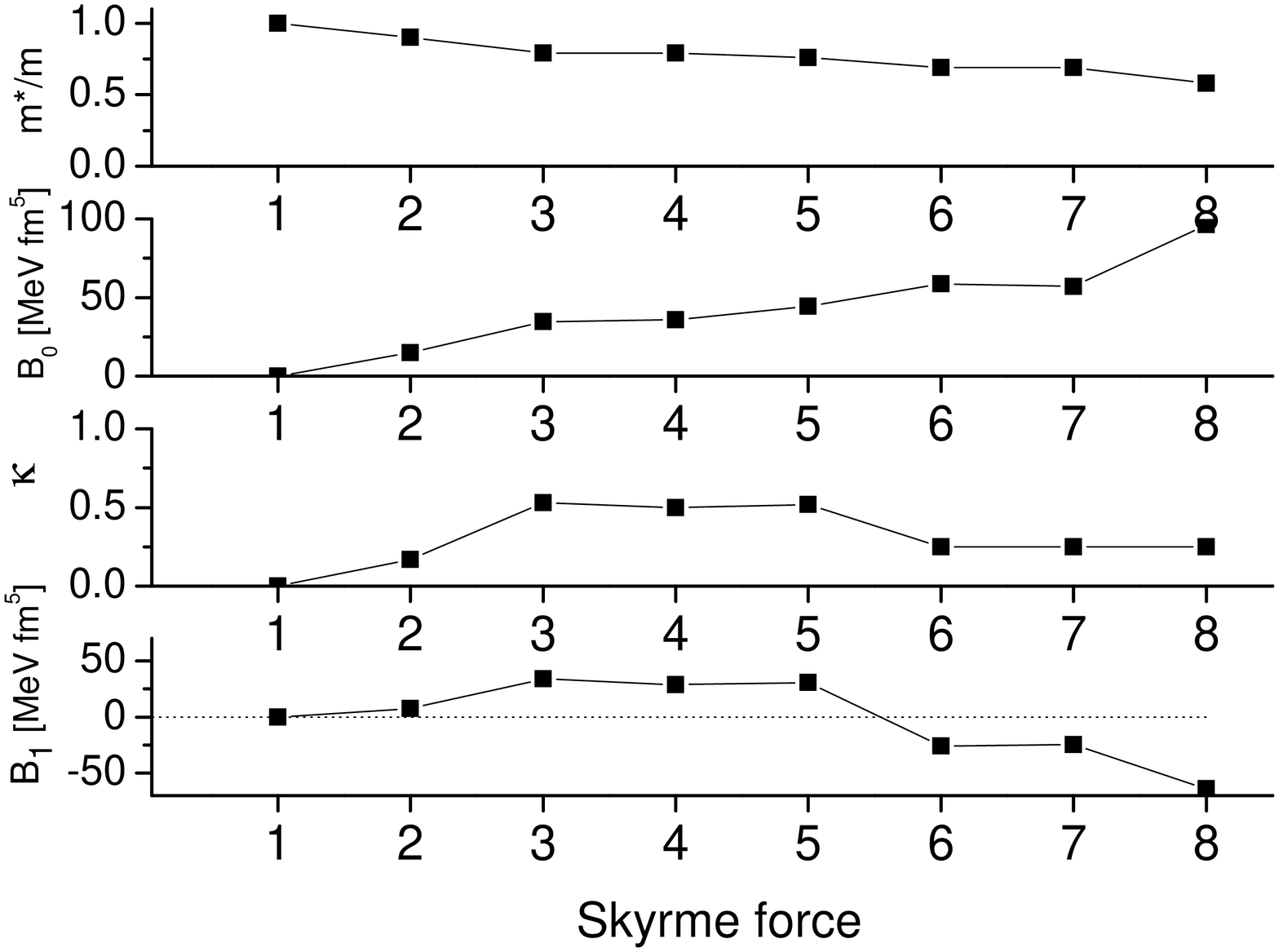}
}
\caption{
Isoscalar $B_0$ and isovector $B_1$ force parameters as well as
related isoscalar effective masses $m*$ and isovector enhancement factors $\kappa$
for the Skyrme forces SkT6 (1), SkO (2), SkM$^*$ (3), SGII (4), SIII (5), SLy4 (6),
SLy6 (7) and SkI3 (8). The dotted line at the bottom plot marks zero value.
}
\end{figure}

The velocity- and spin-dependent densities in the Skyrme functional should influence
energy-weighted sum rules (EWSR) \cite{Boh79,lip_str}. For electric modes,
contributions  of spin-dependent densities $\vec \Im$,  $\vec s$ and
$\vec T$ to EWSR can be neglected (though for Skyrme forces with a low
$m^*$, like SkI3, influence of the spin-orbital density $\vec \Im$ can
be noticeable and reach 7-8$\%$ \cite{Rila}). Instead the contributions of
$\tau$ and $\vec j$ are essential. The former alone results in the effective
mass. However, for isoscalar modes the $\tau$ and $\vec j$ contributions
fully compensate each other \cite{lip_str}. So, if we take $\vec j$ into account,
then EWSR(T=0) again acquires the bare mass $m$ and becomes
\begin{equation}\label{ewsr_0}
EWSR(T=0, \lambda >1)=\frac{(\hbar e)^2}{8\pi m}\lambda (2\lambda +1)^2
A<r^{2\lambda -2}>_A \; .
\end{equation}
Such  compensation is not complete  for isovector modes \cite{lip_str}
and EWSR(T=1,$\lambda \ge 1)$ retains the effective mass. In particular,
\begin{equation}\label{ewsr_1}
EWSR(T=1, \lambda=1)=\frac{(\hbar e)^2}{8\pi m^*_1} 9 \frac{NZ}{A} \; .
\end{equation}

\section{Calculation Scheme}

The study was performed with the representative set of 8 Skyrme forces:
SkT6 \cite{skt6}, SkO \cite{sko}, SIII \cite{sIII}, SGII \cite{sg2},
SkM* \cite{skms}, SLy4 \cite{sly46}, SLy6 \cite{sly46}, SkI3 \cite{ski3}.
The isovector and isoscalar parameters and characteristics of these forces
relevant for our study are exhibited in Fig. 1. We used the SRPA
approach \cite{nest_06,Kva_miori_01,nest_02,ne_les_houches_04,prep_05}
in the approximation of 5 input operators of a different radial dependence.
Four operators were chosen following prescription \cite{nest_02}. For
deformed nuclei, the operator $r^{\lambda +2} Y_{\lambda +2, \mu}$ was added
to take into account the multipole mixing of excitations with the same
projection $\mu^{\pi}$ and parity $\pi$. Direct and exchange Coulomb
contributions to the residual interaction were taken into account. As was shown
in our previous studies \cite{nest_06,nest_02,nest_07}, the proper choice of the
input operators results in sensitivity of the separable residual interaction to
both surface and interior dynamics. Hence high accuracy of the calculations
was achieved already for a few separable terms. Factorization of the residual
interaction results in drastic simplification of RPA calculations, which allows
systematic calculations even for heavy spherical and deformed nuclei.
It should be emphasized that factorization in SRPA is fully self-consistent
and does not lead to additional parameters.

The giant resonances were computed as energy-weighted strength functions
\begin{equation}\label{eq:strength_function}
  S(E\lambda\mu ; E)
 =
  \sum_{\nu}
  E_{\nu} M_{\lambda\mu \nu}^2 \zeta(E - E_{\nu})
\end{equation}
smoothed by the Lorentz function $  \zeta(\omega - \omega_{\nu})  =
\Delta/[2\pi((E - E_{\nu})^2 + (\Delta/2)^2)] $ with the averaging
parameter $\Delta$. In all the calculations $\Delta$= 2 Mev which was found to
be optimal for the comparison with experiment and simulation of various
physical smoothing factors. Further, $M_{\lambda\mu \nu}$ is the matrix element
of $E\lambda\mu$ transition from the ground state to the RPA state $|\nu>$,
$E_{\nu}$ is the RPA eigen-energy. We directly compute the strength
function (\ref{eq:strength_function}) and hence fully avoid determination of
the numerous RPA eigen-states. This additionally reduces the computation time.
SRPA calculations of GR in one nucleus at a familiar laptop take less than one
hour, as compared with weeks while using full RPA methods \cite{Maruhn_PRC_05}.

The E1(T=1) and E2(T=0) resonances were computed with the proton and neutron
effective charges $e_p^{eff}=N/A$, $e_n^{eff}=-Z/A$ and
$e_p^{eff}=e_n^{eff}=1$, respectively. A large configuration space was used.
E.g. in $^{150}$Nd we took into account 247 proton and 307 neutron
single-particle levels ranging from the bottom of the potential
well up to $\sim +20$ MeV. This results in 13000 and 22000
2-quasiparticle configurations for E1(T=1) and E2(T=0) excitations,
respectively, with the energies up to 65 MeV. Then,
EWSR (\ref{ewsr_0}) and (\ref{ewsr_1}) are exhausted by 93-99$\%$ and
95-97$\%$, depending on the force.
%
\begin{figure}[th]
\centerline{
\includegraphics[width=8cm]{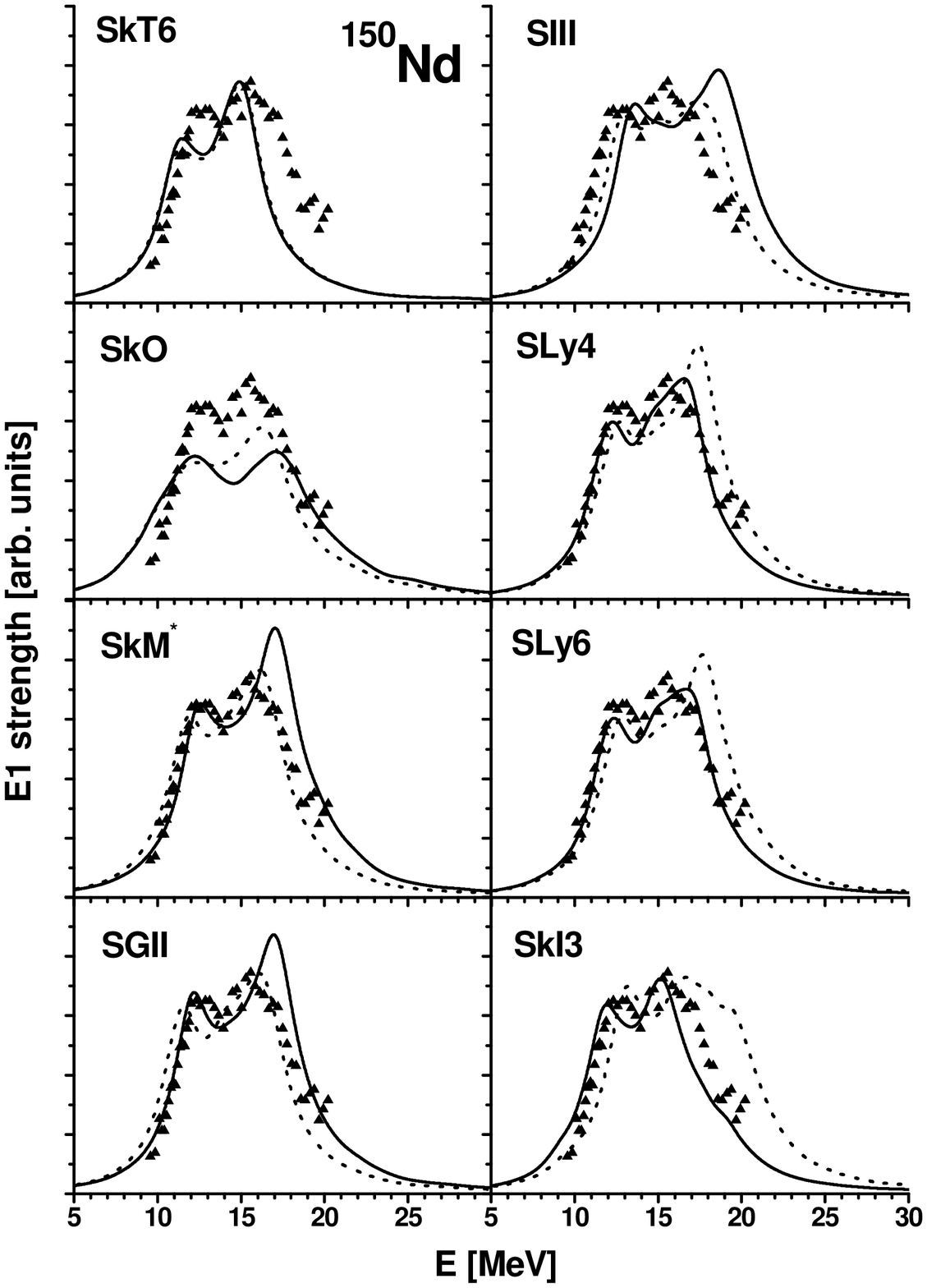}
}
\caption{
E1(T=1) giant resonance in $^{150}$Nd, calculated with the Skyrme
forces SkT6, SkO, SkM$^*$, SGII, SIII, SLy4, SLy6 and SkI3 for the cases
with (solid curve) and without (dotted curve) contribution of the time-odd current.
The strength is smoothed by the Lorentz weight with the averaging  $\Delta$=2 MeV.
The experimental data \protect\cite{nd_exp_carl,nd_exp_berg,nd_exp_varl}
are depicted by triangles.
}
\end{figure}
\begin{figure}
\centerline{
\includegraphics[width=8cm]{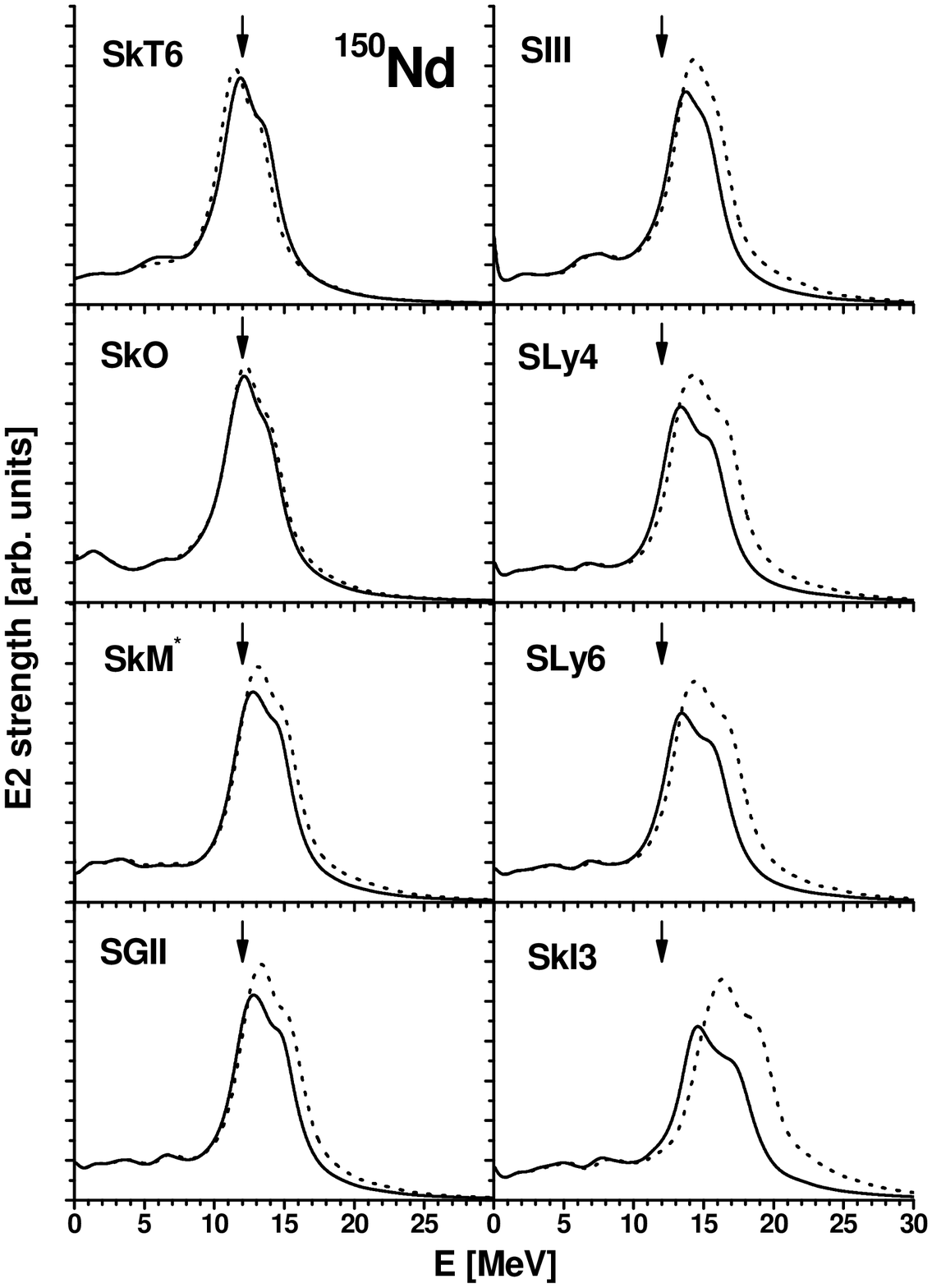}
}
\caption{
The same as in Fig. 1 for E2(T=0) giant resonance. The empirical estimation for
the resonance energy, $E=62 A^{-1/3}$ MeV, is marked the arrows.}
\end{figure}

\section{Results and Discussion}
\label{sec:4}

Results of the SRPA calculations are presented in Figs. 2-7.
In Figs. 2 and 3 the resonances E1(T=1) and E2(T=0) in
$^{150}$Nd are exhibited  for eight Skyrme forces.
It is seen that all the forces give in general acceptable
agreement with the experiment for E1(T=1) resonance, which
confirms ability of SRPA to describe isovector modes. For
E2(T=0) we have no experimental data and so give the empirical estimation.
It is seen that all the forces, for exception of SkT6 and SkO, overestimate
the E2(T=0) energy and the lower the isoscalar effective mass $m^*$
of the force, the more the overestimation. Figs. 2 and 3
demonstrate that no one force can describe simultaneously
E1(T=1) and E2(T=0). This is especially the case for the
forces with $m^*/m < 0.8$.

Figures 2 and 3 compare results obtained with and without
$\vec j$-contribution to the residual interaction.
The careful inspection of the figures shows that $\vec j$-contribution
always improves description of E2(T=0) but not of E1(T=1). Moreover,
the current always results in the down shift for E2(T=0) but may cause
various shifts for E1(T=1). By comparing Figs. 2 and 3 with Fig. 1, one can
see that $\vec j$-effects in E2(T=0) and E1(T=1) are fully determined by
parameters $B_0$ and $B_1$, respectively. The isoscalar parameter $B_0$
is positive for all the forces and steadily increases from SkT6 to SkI3.
Hence we have for E2(T=0) the shift of the same sign for all the forces and
magnitude of the shift rises from SkT6 to SkI3. Instead, the isovector
parameter $B_1$ is zero and small for SkT6 and SkO, positive for SGII, SIII
and SkM* and negative ifor SkT4, SkT6 and SkI3. Hence we have negligible
(SkT6), up (SkM*, SGII, SIII) or down (SkT4, SkT6, SkI3) shifts of E1(T=1).
Magnitude of the shifts obviously correlates with the absolute values of $B_1$.
\begin{figure}[th]
\centerline{
\includegraphics[width=8cm]{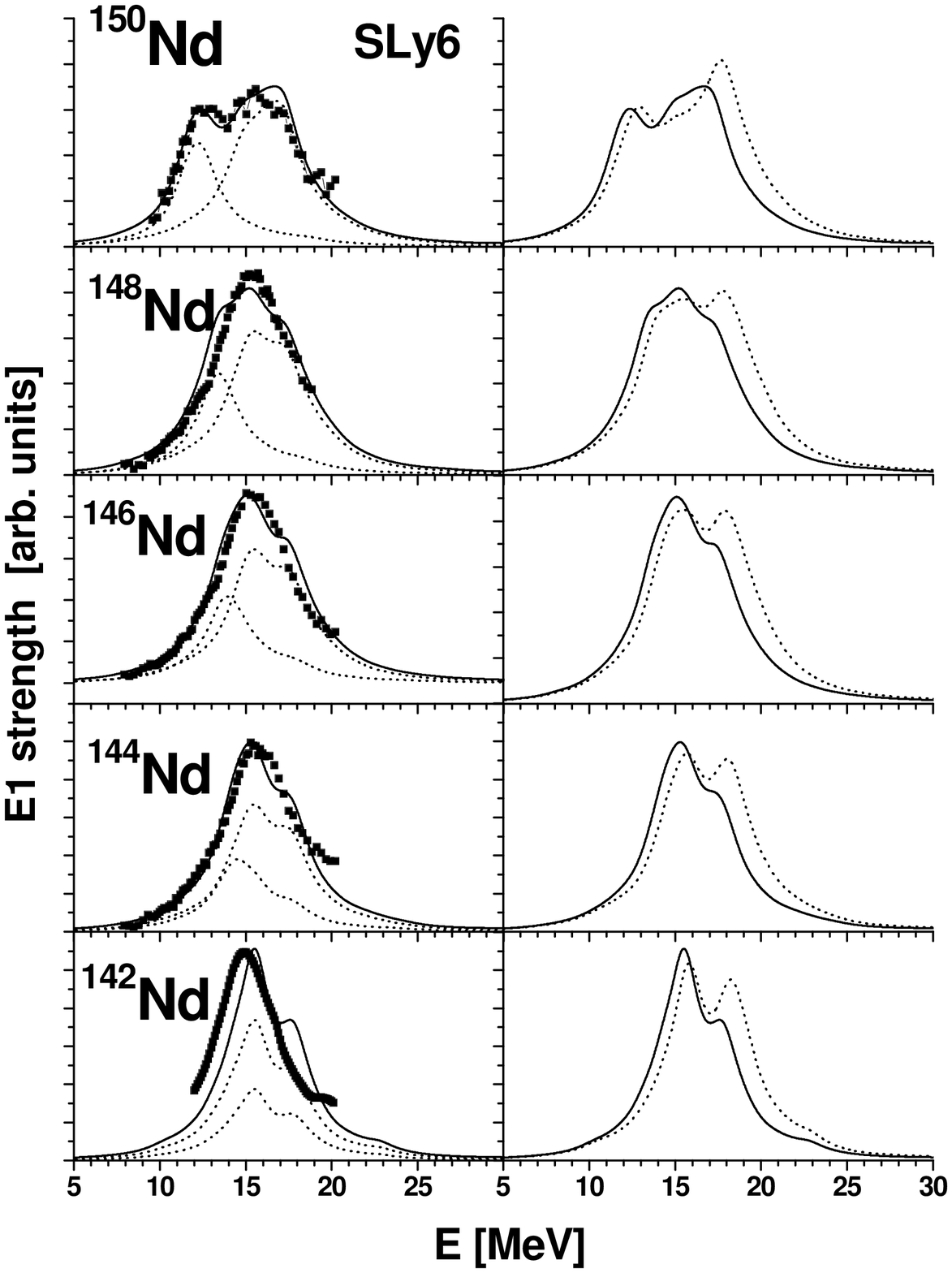}
}
\caption{The dipole giant resonance in Nd isotopes with A=142-150, calculated with
the Skyrme forces SLy6. The left panels exhibit the full strength (solid curve)
calculated with the current contribution. The $\mu=0$ (left small
peak) and $\mu=1$ (right big peak) branches of the resonance are exhibited by dotted
curves. The experimental data
\protect\cite{nd_exp_carl,nd_exp_berg,nd_exp_varl} are depicted by
triangles. The right panels compare the full dipole strengths calculated with
(solid curve) and without (dotted curve) current contribution.
}
\end{figure}
\begin{figure}[th]
\centerline{
\includegraphics[width=8cm]{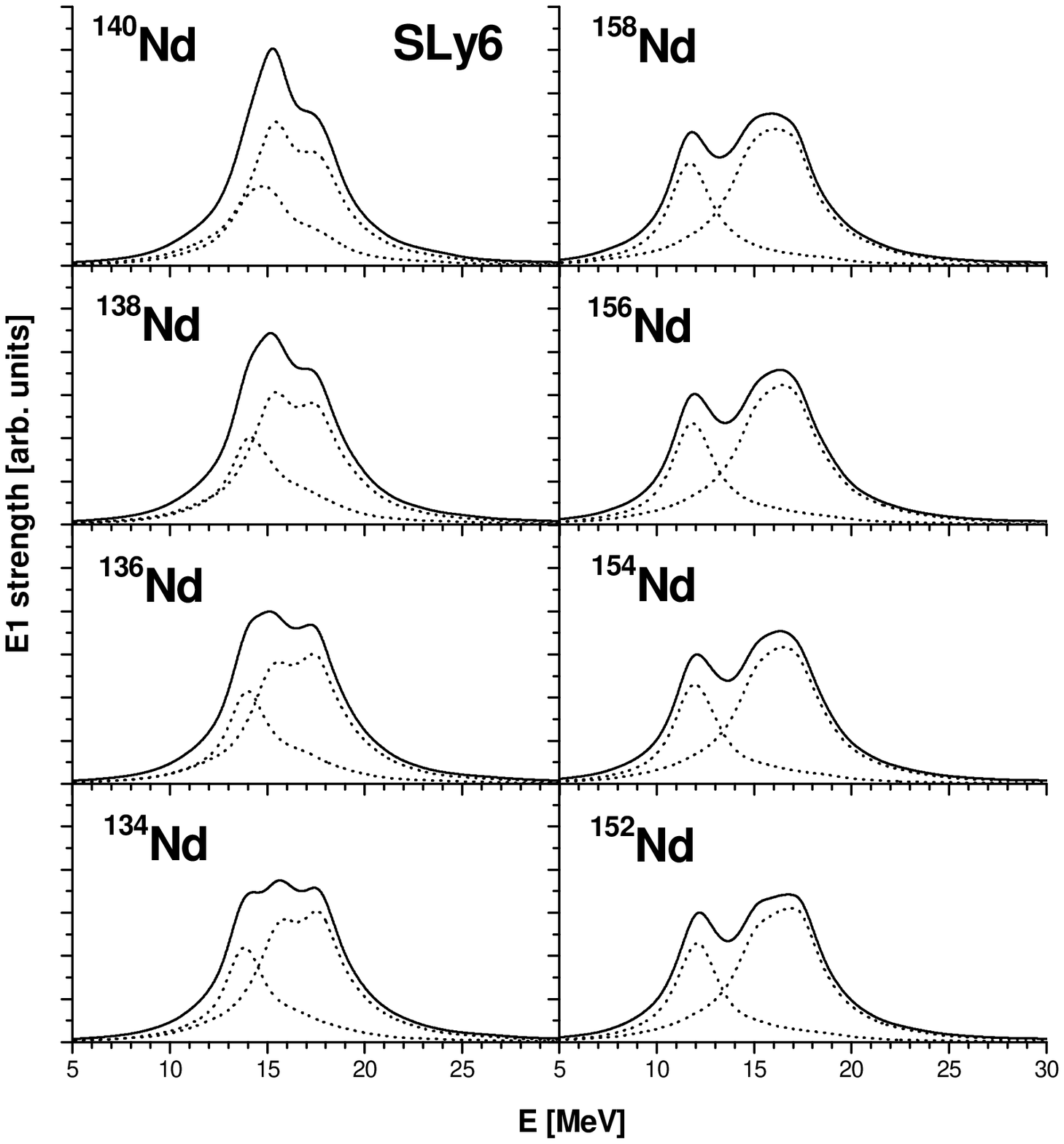}
}
\caption{The dipole giant resonance in Nd isotopes with A=134-140 and 152-158,
calculated for the force SLy6 with the current contribution. The full strength
(solid curve) is supplemented by $\mu=0$ and $\mu=1$ branches (dotted curve).
}
\end{figure}

The above analysis allows to separate the Skyrme forces into 3 groups, depending
on their $B_0$ and $B_1$ parameters. In the first group (SkT6, SkO) these parameters
are zero or small, the effective masses are equal or close to the bare nucleon
mass, and the role of the current density is negligible.  These forces usually
well describe E2(T=0) but fail for E1(T=1).
The second group (SGII, SIII, SkM*) has larger $B_0$ and $B_1$ and therefore
smaller effective masses. What is important, $B_1$ here is positive. These
forces are usually fitted by GR as well (without taking into account time-odd
densities). Hence they provide a reasonable description of GR.  In this case,
adding  $\vec j$-contribution can even worsen the description, as we see in Fig. 2.
These forces violate Galilean invariance and so can hardly be considered as
a robust basis for the further upgrade.
The third group of the forces (SLy4, Sly6, SkI3)
is characterized by low effective masses and, what is important, by the negative
value of $B_1$. These forces (at least SLy4 and Sly6) well describe E1(T=1) but
essentially overestimate the E2(T=0) energy. No one of the three groups is perfect.
More systematic work with Skyrme forces is necessary.

Now we turn to systematic exploration of E1(T=1) resonance in the chain of Nd
isotopes. The calculations were performed with the force SLy6 which, following Fig.2 and our
previous studies \cite{nest_06,nest_07}, promises one of the best descriptions of
E1(T=1). Fig. 4 represents E1(T=1) in the isotopes where we have the
photoabsorption experimental data. The isotopes run from the spherical
semi-magic $^{142}$Nd to the axially deformed $^{150}$Nd discussed above.
The $\mu=0$ and $\mu=1$ brunches of the resonance are exhibited to illustrate
the magnitude and kind (prolate) of the deformation. The figure
demonstrates a remarkable agreement with the experiment for all the nuclei, maybe for
exception of $^{142}$Nd. Following right plots of the figure,
the current contribution is the same for all the isotopes.
\begin{figure}[th]
\centerline{
\includegraphics[width=8cm, angle=-90]{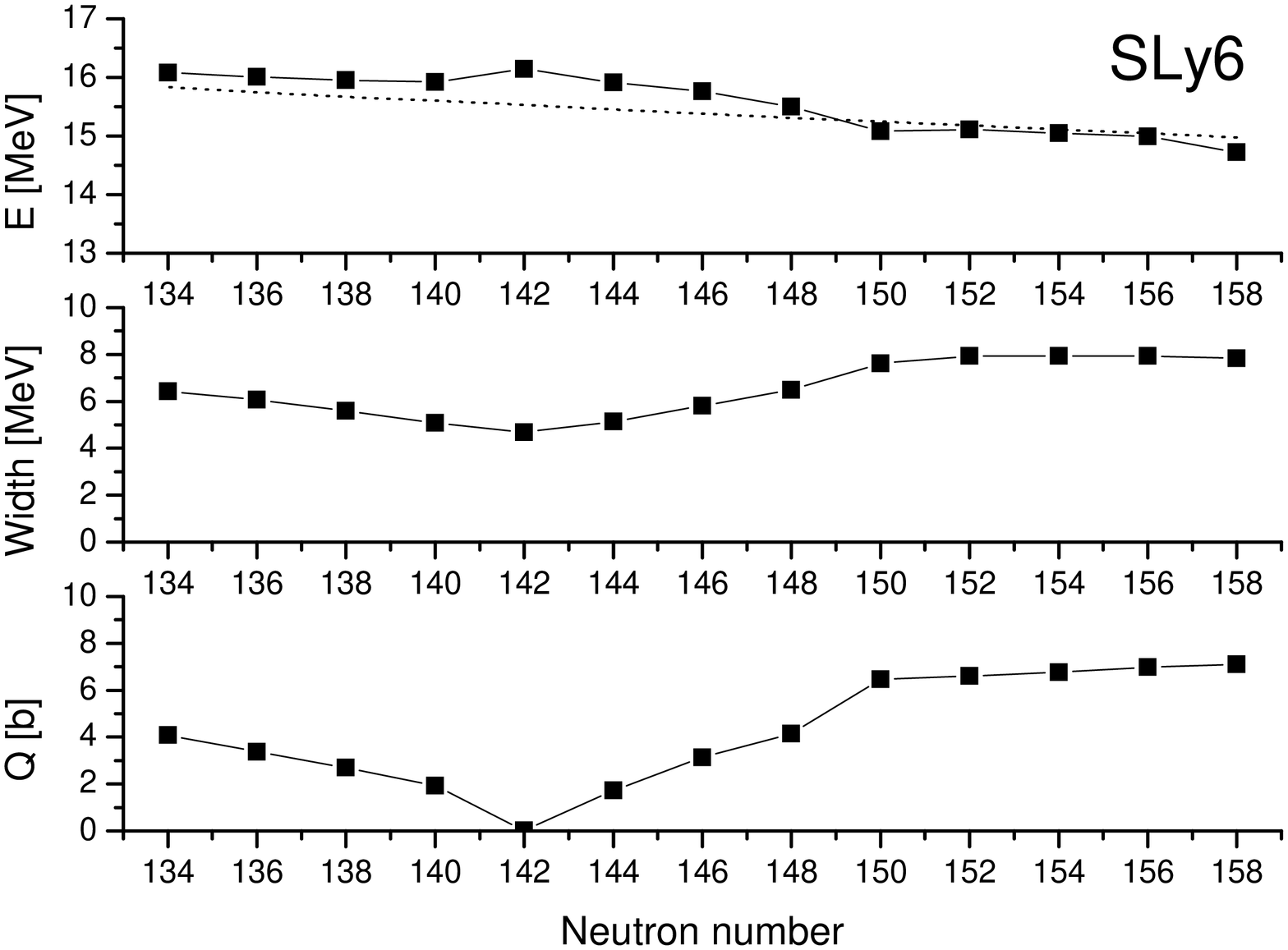}
}
\caption{Energies (upper plot) and widths (middle plot) of the dipole resonance,
calculated (SLy6, with the current contribution) in Nd isotopes with A=134-158.
The bottom plot exhibits the calculated quadrupole moments of the isotopes.
Dotted line at the upper plot gives the empirical estimation $E=81 A^{-1/3}$ MeV
for E1(T=1) energy.
}
\end{figure}
\begin{figure}[th]
\centerline{
\includegraphics[width=10cm]{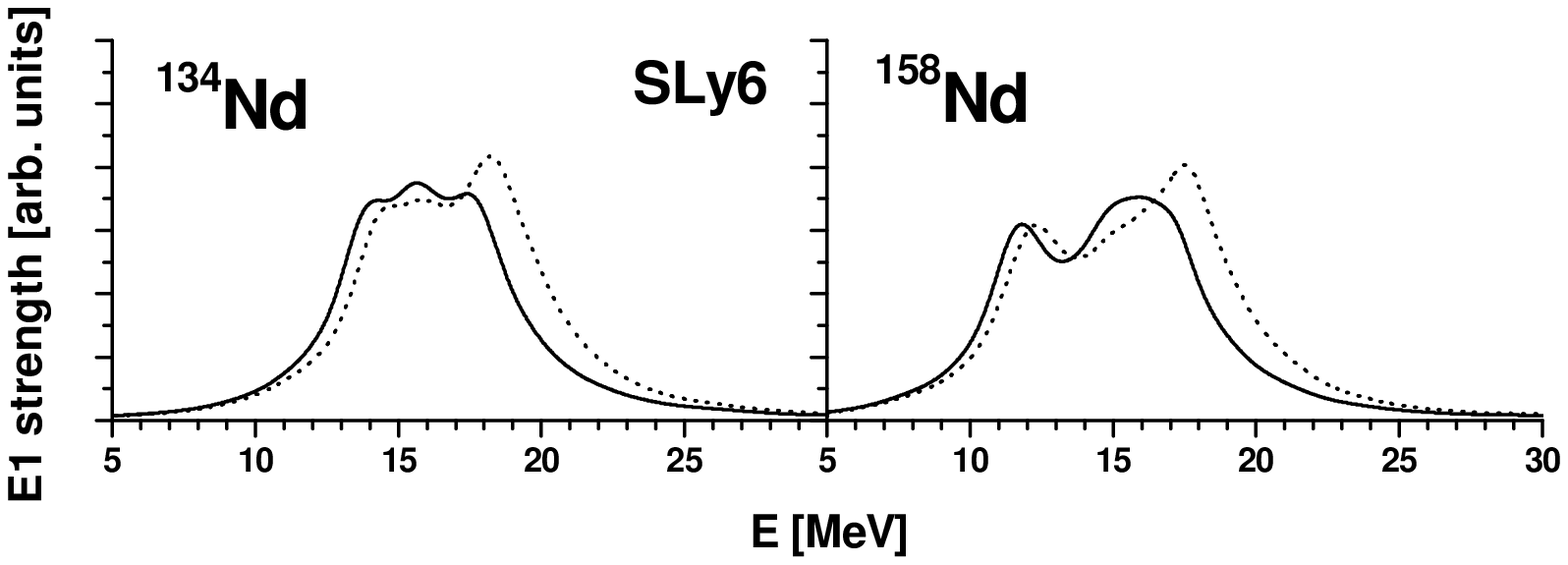}
}
\caption{The dipole resonance calculated with the
force SLy6 for the cases with (solid curve) and without (dotted curve)
contribution of the time-odd current in the isotopes $^{134}$Nd and
in $^{158}$Nd.
}
\end{figure}

In Fig. 5, Nd isotopes with the neutron deficit (A=134-140) and excess (152-158)
are explored. It is interesting that most of the isotopes are deformed
and exhibit prolate shape. The cartesian quadrupole moments for the isotopes
are given in Fig. 6. This figure also shows E1(T=1) centroids and
widths. The latter obviously correlate with the magnitude of the deformation.
The resonance energies generally agree with estimation $E=81 A^{-1/3}$ MeV.

Finally, Fig. 7 demonstrates that the current density contribution to E1(T=1)
ends of the isotope chain is the same as in the middle of the chain (compare
Figs. 4 and 7). So $\vec j$-impact is fully determined by the Skyrme force and
does not depend on the shape and neutron number of the isotope. Since
$\vec j$-impact is related with the effective masses, this observation sends a
promising message that results for time-odd densities and effective masses
obtained for standard nuclei may be relevant for exotic areas as well.

\section{Conclusions}

The effect of time-odd densities was discussed and examined for the particular case
of the current density. The contribution of this density to E1(T=1) and E2(T=0) giant
resonance was explored in the chain of Nd isotopes for a representative set of 8 Skyrme
forces. The current impact was found to be generally strong and fully determined by the
isoscalar and isovector parameters $B_0$ and $B_1$  of the Skyrme forces, responsible
for the effective masses. The results obtained allow to classify the Skyrme forces into
3 groups, depending on the magnitude and sign of $B_1$. Our study show that time-odd
densities represent a principle  ingredient of Skyrme forces, responsible for the
Galilean invariance. They are closely related with the effective masses and
significantly influence nuclear dynamics. So, time-odd densities can be an important
factor for further upgrade and unification of Skyrme forces.

The calculations show that the current impact is fully determined by the Skyrme force
and does not depend on the shape or neutron number of the isotope. In other words, the
impact is the same for standard and exotic nuclei. This message hints an
interesting possibility to explore the time-odd density effects for standard nuclei
(where there are  experimental data) and then to transfer the results to exotic areas.

\section*{Acknowledgments}

The work was partly supported  by DFG grant RE 322/11-1 and
Heisenberg-Landau (Germany-BLTP JINR) grants for 2006 and 2007 years.
W.K. and P.-G.R. are grateful for the BMBF support under contracts
06 DD 139D and 06 ER 808. This work is also a part of the research
plan MSM 0021620834 supported by the Ministry of Education of the
Czech Republic. It was partly funded by Czech grant agency
(grant No. 202/06/0363) and grant agency of
Charles University in Prague (grant No. 222/2006/B-FYZ/MFF).

\end{document}